\documentclass[ reprint, amssymb, amsmath, pre,cha]{revtex4-1}

\usepackage[colorlinks=true,linkcolor=blue]{hyperref}%
\usepackage{graphicx}
\usepackage{amsmath,accents}
\usepackage{blkarray}
\usepackage{booktabs}

\newcommand{\la}{\ensuremath{\langle}}
\newcommand{\ra}{\ensuremath{\rangle}}

\newcommand{\vc}{\mathbf}

\usepackage{color}
\usepackage{amsmath,amssymb}
\usepackage{graphicx}
\usepackage{empheq}
\usepackage{mathrsfs}

\begin{document}

\title{Viscosity of the magnetized  strongly coupled one-component plasma}
\author{Brett Scheiner }
\affiliation{Los Alamos National Laboratory, Los Alamos, New Mexico, 87545}
\email{bss@lanl.gov}

\author{Scott D. Baalrud}
\affiliation{Department of Physics and Astronomy, University of Iowa, Iowa City, Iowa 52242}

\begin{abstract}

The viscosity tensor of the magnetized one-component plasma, consisting of five independent shear viscosity coefficients, a bulk viscosity coefficient, and a cross coefficient, is computed using equilibrium molecular dynamics simulations and the Green-Kubo relations. 
A broad range of Coulomb coupling and magnetization strength conditions are studied. 
Magnetization is found to strongly influence the shear viscosity coefficients when the gyrofrequency exceeds the Coulomb collision frequency. 
Three regimes are identified as the Coulomb coupling strength and magnetization strength are varied. 
The Green-Kubo relations are used to separate kinetic and potential energy contributions to each viscosity coefficient, showing how each contribution depends upon the magnetization strength. 
The shear viscosity coefficient associated with the component of the stress tensor parallel to the magnetic field, and the two coefficients associated with the component perpendicular to the magnetic field, are all found to merge to a common value at strong Coulomb coupling. 

\end{abstract}

\maketitle

\section{Introduction}

Viscosity is a material property that determines how a plasma responds to shear stress (shear viscosity) or compression (bulk viscosity). 
It must be well characterized in order to accurately model flow profiles and viscous heating rates. 
It contributes to dimensionless parameters, such as the Reynold's number, Prandtl number and magnetic Prandtl number, that characterize a wide range of important processes in plasmas, including turbulence~\cite{DavidovitsPRL2016}, magnetic reconnection~\cite{ParkPOP1984,ComissoJPP2015}, and dynamo amplification of magnetic fields~\cite{SchekochihinPRL2004,TzeferacosNC2018}. 
Current understanding of the microscopic origin of shear viscosity is largely based upon the Braginskii transport theory~\cite{1958JETP....6..358B,1965RvPP....1..205B}, which is a Chapman-Enskog solution of the Boltzmann kinetic equation for a plasma~\cite{chapman1939mathematical}. 
This theory applies to conditions in which each species of the plasma is both weakly coupled ($\Gamma \ll 1$) and weakly magnetized ($\beta \ll 1$). 
Here, coupling strength is characterized by the Coulomb coupling parameter
\begin{equation}
\label{eq:gamma}
    \Gamma \equiv \frac{e^2/a}{k_BT}
\end{equation}
where $e$ is the electronic charge, $a=(3/4\pi n)^{1/3}$ is the average interparticle spacing, and $T$ is the temperature. 
Magnetization strength is characterized by the magentization parameter 
\begin{equation}
\label{eq:beta}
    \beta \equiv \frac{\omega_c}{\omega_p}
\end{equation}
where $\omega_c = e|B|/m$ is the gyrofrequency and $\omega_p = \sqrt{e^2 n/\epsilon_o m}$ is the plasma frequency. 

Although plasmas are commonly weakly coupled and weakly magnetized, by these measures, there are also many examples in which the Coulomb coupling strength and the magnetization strength can have moderate-to-large values ($\Gamma \gtrsim 0.1$ or $\beta \gtrsim 0.1$). 
These include trapped non-neutral plasmas~\cite{DubinRMP1999,KrieselPRL2001}, ultracold neutral plasmas~\cite{ZhangPRL2008}, as well as dense plasmas created in inertial confinement fusion experiments~\cite{GomezPRL2014}, high energy density plasma experiments~\cite{ChittendenPRL2007}, and those found in nature, such as dense stars~\cite{UzdenskyApJ2014} and giant planets~\cite{FortneyPOP2009}. 
There is little understanding of how the combined effects of Coulomb coupling and magnetization strength influence viscosity. 

This paper presents first-principles computations of the viscosity of the one-component plasma (OCP) at conditions ranging from moderate to strong coupling ($\Gamma = 1, 10$ and $100$) and weak to moderate magnetization ($\beta = 0.01-2$) using molecular dynamics (MD) simulations. 
The magnetized OCP is a model system in which only one species is dynamical, but which is assumed to evolve in the presence of a non-interacting and non-polarizable neutralizing background~\cite{bausPR1980}. 
It is convenient for studying the fundamental physics associated with coupling and magnetization strength because it is completely characterized by the two dimensionless parameters of Eqs.~(\ref{eq:gamma}) and (\ref{eq:beta}). 
Previous work has explored diffusion~\cite{OttPRL2011,BaalrudPRE2017}, thermal conduction~\cite{OttPRE2015}, temperature anisotropy relaxation~\cite{BaalrudPRE2017}, and friction~\cite{Bernstein2020} of the magnetized OCP. The viscosity of a related system, the 2D magnetized Yukawa OCP, has also been studied~\cite{PhysRevE.96.053208}. While these results are relevant to dusty plasma experiments~\cite{PhysRevLett.93.155004}, they do not translate to inform the behavior the viscosity tensor in three-dimensional systems.

Although the OCP is a model system, certain properties are also quantitatively applicable to real plasmas. 
Viscosity is one of these properties. 
Because momentum transfer in an electron-ion plasma is predominately determined by the more massive ion species, the total plasma viscosity is usually associated with the ion contribution alone. 
When electron dynamics are negligible, the ion viscosity coefficients can be obtained from the OCP~\cite{PhysRevA.34.4163}.

\begin{figure}
\includegraphics[width=1\columnwidth]{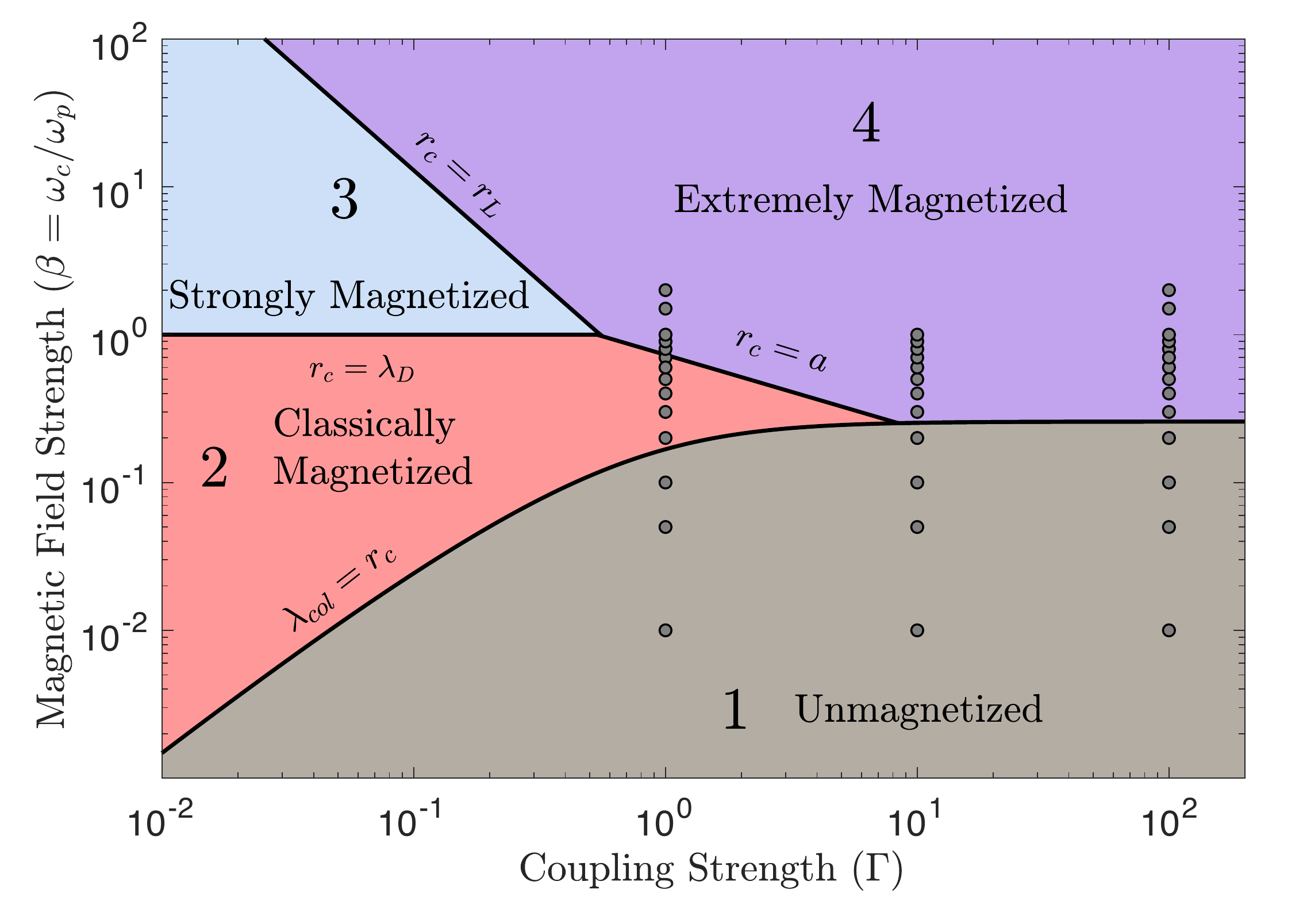}
\label{fg:gammabeta}
\caption{Predicted regimes in which transport coefficients are determined by different microphysical processes in terms of Coulomb coupling and magnetization strength. Circles indicate the conditions of the MD simulations. }
\end{figure}

The MD simulation results reveal a number of interesting features. 
One is that scaling laws of the various shear viscosity coefficients transition between regimes at boundaries in coupling-magnetization parameter-space that are defined by comparing the gyroradius $r_c = \sqrt{k_BT/m}/\omega_c$ to either the Coulomb collision mean free path $\lambda_{\textrm{col}}$ (as defined in~\cite{BaalrudPRE2017}), Debye length $\lambda_D = \sqrt{\epsilon_o k_BT/e^2n}$, or the minimum interaction scale length, which is characterized by the minimum of the thermal distance of closest approach (i.e., Landau length) times $\sqrt{2}$, $r_L = \sqrt{2}e^2/k_BT$, the average interparticle spacing, $a$, or the Coulomb collision mean free path. 
These regime boundaries, which were recently proposed in~\cite{BaalrudPRE2017}, are shown in Fig.~\ref{fg:gammabeta}. 
MD simulation data are obtained at conditions that access regions 1, 2 and 4 in this proposed parameter space, showing that fundamental transitions in the scaling of shear viscosity coefficients with $\beta$ occur as these boundaries are crossed. 

In a magnetized plasma, viscosity is described by a fourth-rank tensor that nominally consists of 81 components. 
However, symmetry associated with the magnetic field being straight and uniform, as well as the Onsager reciprocal relations, reduces this to a tensor described by five independent shear viscosity coefficients, one bulk viscosity coefficient, and one coefficient associated with coupling between bulk and shear viscosity~\cite{1954Phy....21..355H,1969net..book.....D}. 
In the weakly magnetized regime (region 2), Braginskii theory predicts that the shear viscosity coefficient ($\eta_o^\textrm{B}$) associated with shear stress parallel to the magnetic field $\Pi_\parallel$ is independent of the magnetic field, the two coefficients ($\eta_1^{\textrm{B}}$ and $\eta_2^{\textrm{B}}$) associated with shear stress perpendicular to the magnetic field $\Pi_\perp$ are proportional to $\beta^{-2}$, and the two coefficients ($\eta_3^{\textrm{B}}$ and $\eta_4^{\textrm{B}}$) associated with shear stress in the transverse direction $\Pi_\wedge$ are proportional to $\beta^{-1}$~\cite{1958JETP....6..358B,1965RvPP....1..205B,1983nrl..reptQ....B}. 
It also predicts that the bulk viscosity coefficient ($\mu_v$) and the coefficient associated with coupling of shear and bulk viscosity ($\zeta$) are both zero. 
Although the lowest coupling strength simulated in our work was $\Gamma = 1$, which accessed only a small region of the weakly magnetized regime, the results obtained are consistent with these predictions for $\eta_o^{\textrm{B}}$, $\eta_1^{\textrm{B}}$ and $\eta_2^{\textrm{B}}$. 
The simulations are also consistent with $\zeta$ and $\mu_v$ being zero, but are unable to resolve $\eta_3^{\textrm{B}}$ and $\eta_4^{\textrm{B}}$ due to the achievable level of numerical accuracy. 

Qualitatively new behavior is observed in the transition from either regions 1 or 2, to 4. 
At the lower coupling strength values ($\Gamma =1$ and $10$), the $\eta_o^\textrm{B}$ coefficient is observed to become dependent on the magnetic field strength, scaling as a positive power of $\beta$, while the $\eta_1^{\textrm{B}}$ and $\eta_2^{\textrm{B}}$ coefficients are found to transition from scaling as a negative power of $\beta$ to become nearly independent of $\beta$, or possibly as a slightly positive power of $\beta$ in region 4. 
In the strongly coupled case, $\Gamma = 100$, the $\eta_o^\textrm{B}$, $\eta_1^{\textrm{B}}$ and $\eta_2^{\textrm{B}}$ coefficients are all observed to merge to a common value. 
This common value is independent of the magnetic field strength in region 1, and scales as a positive power of $\beta$ in the transition to region 4. 
In all cases considered, all of the remaining coefficients, $\eta_3^{\textrm{B}}$, $\eta_4^{\textrm{B}}$, $\zeta$ and $\mu_v$, are consistent with zero to within the accuracy of the simulations; although they likely have finite values they are smaller than the other viscosity coefficients and were unable to be resolved. 

These calculations are based upon the Green-Kubo relations, which in addition to the total viscosity coefficients also provides information about their physical origin. 
In particular, the shear-stress autocorrelation function can be split into a kinetic component that depends only on the particle momenta, and a potential component that depends on the particle positions. 
Cross terms are also present, but are small. 
Previous MD simulations of the unmagnetized OCP~\cite{BernuPRA1978,BasteaPRE2005,DonkoPRE2008,DaligaultPRE2014} have established that the kinetic component is dominant when $\Gamma \lesssim 17$, that the potential component is dominant when $\Gamma \gtrsim 17$, and that the total shear viscosity coefficient has a minimum value at this transition $\Gamma \approx 17$. 
We find that in the magnetized case, the $\eta_o^\textrm{B}$, $\eta_1^{\textrm{B}}$ and $\eta_2^{\textrm{B}}$ all converge to the result obtained in previous simulations in the unmagnetized regime, as expected. 
Furthermore, it is shown that the transition from predominantly kinetic to predominately potential contributions depends on $\beta$ as well as $\Gamma$, and it differs for each coefficient. 
Both the kinetic and potential contributions are influenced by the magnetic field. 

This paper is organized as follows: Sec.~\ref{sec:visc} provides an overview of different schemes for describing shear viscosity of a magnetized plasma. 
Sec.~\ref{sec:setup} describes the simulation technique, Sec.~\ref{sec:calc} the Green-Kubo relations for the viscosity coefficients, Sec.~\ref{sec:convergence} an analysis of the conditions for convergence, and Sec.~\ref{sec:results} the results of the calculation. 
A discussion of the results is provided in Sec.~\ref{sec:discussion}, and concluding comments in Sec.~\ref{sec:conclusion}.

\section{Viscosity of a magnetized fluid~\label{sec:visc}}

In a magnetized fluid, the viscous stress tensor $\vc{\Pi}$ and the rate-of-strain tensor $\vc{W}$ are rank-2 tensors due to the anisotropy introduced by the magnetic field. The shear viscosity tensor that provides a linear relation between these quantities is a rank-4 tensor   
\begin{equation}
\label{eq:Pi_ab}
\Pi_{\alpha\beta}=-L_{\alpha\beta\gamma\delta}W_{\gamma\delta},
\end{equation}
in which the Cartesian indices $\alpha$, $\beta$, $\gamma$, $\delta$ run from 1 to 3, and 
\begin{equation}
\vc{W} \equiv \frac{1}{2}[\nabla \vc{V} + (\nabla \vc{V})^T] .
\end{equation}

A Cartesian rank-4 tensor has 81 components. 
However, general symmetry arguments of non-equilibrium thermodynamics, the Onsager reciprocal relations, and the assumption of a straight uniform magnetic field, can be used to show that the shear viscosity tensor can be put into a more intuitive and tractable form with just 7 independent coefficients~\cite{1954Phy....21..355H,1969net..book.....D}.
Here, we summarize these symmetry arguments and how the tensor in Eq.~(\ref{eq:Pi_ab}) can be related to the form of the shear viscosity tensor that is commonly used in plasma physics~\cite{1983nrl..reptQ....B}.

In Eq.~(\ref{eq:Pi_ab}), $\vc{W}$ can split into a sum of its trace and traceless components: 
\begin{equation}
\label{eq:grad_v}
\vc{W}=\frac{1}{3}(\nabla\cdot\vc{V})\vc{I}+\vc{S}
\end{equation}
where $\vc{S}$ is the rate-of-shear tensor. The same can also be done for the viscous stress tensor, which can be split into the bulk and shear viscous stress as
\begin{equation}
\label{eq:pi}
\vc{\Pi}=\frac{\pi_\textrm{T}}{3}\vc{I}+\mathring{\vc{\Pi}},
\end{equation}
respectively. Here, the trace of the viscous stress tensor is  
\begin{equation}
\pi_\textrm{T} =\sum_{\alpha} \Pi_{\alpha\alpha}.
\end{equation}
Since the stress tensor is symmetric, its components can be written in terms of a rank-2 tensor with indices running from 1 to 6 using Voigt notation~\cite{voigt1910lehrbuch}
\begin{equation}
-\pi_i=\sum_{k=1}^6L_{ik}w_k,
\end{equation}
where $\pi_i$ are the elements of $\vc{\Pi}$ and the indices are shorthand as follows: $1=xx$, $2=yy$, $3=zz$, $4=yz$, $5=xz$, and $6=xy$. The elements of $\vc{W}$ are $w_1=W_{xx}$, $w_2=W_{yy}$, $w_3=W_{zz}$, $w_4=2W_{yz}=2W_{zy}$, $w_5=2W_{xz}=2W_{zx}$, and  $w_6=2W_{xy}=2W_{yx}$.
The factor of two in the last three components appears because they appear twice (e.g. $xy$ and $yx$ for $w_6$). In this notation the 81-component Cartesian rank-4 tensor $L_{\alpha\beta\gamma\delta}$ can be reduced to a 36-component rank-2 tensor $L_{ik}$: 
\begin{equation}
\begin{blockarray}{ccccccc}
& w_1&w_2&w_3&w_4&w_5&w_6\\
\begin{block}{l(cccccc)}
  -\pi_1 \ \ \  & L_{11} &  L_{12} &  L_{13} &  L_{14} &  L_{15} &  L_{16}  \\
  -\pi_2 \  & L_{21} &  L_{22} &  L_{23} &  L_{24} &  L_{25} &  L_{26}  \\
  -\pi_3 \  & L_{31} &  L_{32} &  L_{33} &  L_{34} &  L_{35} &  L_{36}  \\
  -\pi_4 \  & L_{41} &  L_{42} &  L_{43} &  L_{44} &  L_{45} &  L_{46}  \\
  -\pi_5 \  & L_{51} &  L_{52} &  L_{53} &  L_{54} &  L_{55} &  L_{56}  \\
    -\pi_6 \  & L_{61} &  L_{62} &  L_{63} &  L_{64} &  L_{65} &  L_{66}  \\
\end{block}\label{eq:lmatrix}
\end{blockarray} \ \ \ .
\end{equation}
The formulation of a rank-4 tensor in this manner is known as Voigt notation. Appendix A summarizes coordinate rotation properties of Cartesian rank-4 tensors expressed in this notation.

The form of $L_{ik}$ can be simplified further by assuming a uniform magnetic field~\cite{1954Phy....21..355H}, chosen here to be parallel to the $z$-axis. 
A rotation about the $z$-axis should leave the elements of $L_{ik}$ invariant. After a 180$^\circ$ rotation about the $z$-axis, the following elements pick up a negative sign and therefore must be zero: $L_{14}$, $L_{15}$, $L_{24}$, $L_{25}$, $L_{34}$, $L_{35}$, $L_{41}$, $L_{42}$, $L_{43}$, $L_{46}$, $L_{51}$, $L_{52}$, $L_{53}$, $L_{56}$, $L_{64}$, $L_{65}$. 
The same conclusion can be drawn for $L_{63}$ and $L_{36}$ after an infinitesimal rotation about $z$. After a 180$^\circ$ rotation about the $x$ axis, the system obeys the parity relation $L_{ik}(\vc{B})=(-1)^nL_{ik}(-\vc{B})$, where $n$ is the number of times $x$ appears in the indices $ik$. The same relation also holds for a 180$^\circ$ rotation about $y$. 
This leads to the conclusion that $L_{11}$, $L_{22}$, $L_{33}$, $L_{12}$, $L_{21}$, $L_{13}$, $L_{31}$, $L_{32}$, $L_{23}$, $L_{44}$, $L_{55}$, and $L_{66}$ are even functions of $\vc{B}$ and that $L_{16}$, $L_{26}$, $L_{45}$, $L_{54}$, $L_{61}$, and $L_{62}$ are odd functions of $\vc{B}$. 
Additional simplifications are made by noting the similarity between the $x$ and $y$ coordinate axes: $L_{11}=L_{22}$, $L_{44}=L_{55}$, $L_{31}=L_{32}$, $L_{13}=L_{23}$.

The microscopic reversibility of the system under a sign change of time and the magnetic field direction also plays a role in the form of the shear viscosity tensor. The Onsager reciprocal relations relate components of the viscosity tensor on the basis of symmetries resulting from this reversibility. For a magnetized plasma where the particles interact via the Lorentz force, the coefficients are related by $L_{ik}(\vc{B})=L_{ki}(-\vc{B})$~\cite{RevModPhys.17.343}. Using the fact that $L_{13}$ and $L_{23}$ are even functions of $\vc{B}$, this leads to the relation $L_{13}=L_{31}$ and $L_{23}=L_{32}$.
With these simplifications, the shear viscosity tensor can be expressed as
\begin{equation}
\label{eq:pi_epsilon}
\begin{blockarray}{ccccccc}
& w_1&w_2&w_3&w_4&w_5&w_6\\
\begin{block}{l(cccccc)}
  -\pi_1 \ \ \  & L_{11} &  L_{12} &  L_{13} & 0  & 0 &  L_{16}  \\
  -\pi_2 \  & L_{12} &  L_{11} &  L_{13} & 0 & 0  &  -L_{16}  \\
  -\pi_3 \  & L_{13} &  L_{13} &  L_{33} & 0 & 0  & 0 \\
  -\pi_4 \  &0 & 0  & 0 &  L_{44} &  L_{45} & 0  \\
  -\pi_5 \  &0 &0 & 0  &  -L_{45} &  L_{44} & 0  \\
    -\pi_6 \  & -L_{16} &  L_{16}& 0 & 0 & 0 &  L_{66}  \\
\end{block}
\end{blockarray} \ \ \ .
\end{equation}

\

\

Equation~(\ref{eq:pi_epsilon}) can be split into components associated with the shear viscosity and the bulk viscosity by following the method of Hooyman, DeGroot, and Mazur~\cite{1954Phy....21..355H,1969net..book.....D}. 
This is done in two steps. The first is to separate the bulk viscosity by splitting $\vc{W}$ and $\vc{\Pi}$ into their traces and traceless parts given by Eqs.~(\ref{eq:grad_v}) and (\ref{eq:pi}). 
In terms of Voigt notation, $\vc{S}$ in Eq.~(\ref{eq:grad_v}) has components $s_i=w_i-\nabla\cdot\vc{V}/3$ for $i=1$ to 3 and the traceless components of Eq.~(\ref{eq:pi}) are $\mathring{\pi}_i=\pi_i-\pi_\textrm{T}/3$, also for $i=1$ to 3. 
Next, by writing  $\mathring{\pi}_1$, $\mathring{\pi}_2$, $\mathring{\pi}_3$, $\pi_4$, $\pi_5$, $\pi_6$, and $\pi_\textrm{T}$ in terms of the corresponding $L_{ik}$ components of $s_1$,  $s_2$, $s_3$, $w_4$, $w_5$, $w_6$ and $\frac{1}{3}(\nabla\cdot\vc{V})$ and making use of the relation $s_1+s_2+s_3\equiv0$ to write $s_3 = -s_1 - s_2$, Eq.~(\ref{eq:pi_epsilon}) can be written as
\begin{widetext}
\begin{equation}
\label{eq:Mazur_pi}
\begin{blockarray}{cccccccc}
&s_1&s_2&s_3&w_4&w_5&w_6&\frac{\nabla\cdot\vc{V}}{3}\\
\begin{block}{l(ccccccc)}
  -\mathring{\pi}_1 \ \ \  &2\mu_2 &2(\mu_1-\mu_2) &0  & 0  & 0 &\eta_1   &-\zeta \\
  -\mathring{\pi}_2 \  & 2(\mu_1-\mu_2) & 2\mu_2 & 0 & 0 & 0  &  -\eta_1  &-\zeta\\
  -\mathring{\pi}_3 \  &0 &  0 & 2\mu_1 & 0 & 0  & 0 &2\zeta\\
  -\pi_4 \  &0 & 0  & 0 & \mu_3  & \eta_2  & 0&  0\\
  -\pi_5 \  & 0&   0&  0 & -\eta_2 & \mu_3 & 0&  0\\
    -\pi_6 \  &-\eta_1  & \eta_1 & 0 & 0 &0 &  2\mu_2-\mu_1 & 0\\
     -\pi_\textrm{T} \  & -\zeta & -\zeta& 2\zeta & 0 & 0&0  &9\mu_v , \\
\end{block}
\end{blockarray} 
\end{equation}
\end{widetext}
where 
\begin{subequations}
\begin{eqnarray}
\mu_1&=&\frac{1}{6}(2L_{33}-4L_{13}+L_{11}+L_{12}),\label{HDMmu1}\\
\mu_2&=&\frac{1}{6}(2L_{11}-L_{12}-2L_{13}+L_{33}),\label{HDMmu2}\\
\mu_3&=&L_{44},\\
\eta_1&=&L_{16},\\
\eta_2&=&L_{45},\\
\mu_v&=&\frac{1}{9}(2L_{11}+2L_{12}+4L_{13}+L_{33}),\label{HDMBulk}\\
\zeta&=&\frac{1}{3}(L_{13}+L_{33}-L_{11}-L_{12})\label{HDMzeta}.
\end{eqnarray}
\end{subequations}
This form, commonly used in non-equilibrium thermodynamics, expresses the viscosity tensor in terms of the five shear viscosity coefficients $\mu_1$, $\mu_2$, $\mu_3$, $\eta_1$ and $\eta_2$, the bulk viscosity coefficient $\mu_v$ and a ``cross coefficient'' $\zeta$. 

In plasma physics, it is more common to write the shear viscosity coefficients in the form expressed in Braginskii's review~\cite{1958JETP....6..358B,1965RvPP....1..205B}. 
This makes use of the result of the Chapman-Enskog solution of the plasma kinetic equation, which predicts that the bulk viscosity and cross coefficients are zero in a weakly coupled plasma ($\zeta = 0$ and $\mu_v = 0$). 
In this limit, the last equation in (\ref{eq:Mazur_pi}) is $\pi_\textrm{T} = 0$, which from the definition in Eq.~(\ref{eq:pi}) implies that $\vc{\Pi} = \mathring{\vc{\Pi}}$. 
The result can be expressed as only six equations describing the shear viscosity components Eq.~(\ref{eq:Mazur_pi}) as~\footnote{In the notation of the NRL plasma formulary $\vc{W}\equiv2\vc{S}$, with a different definition for $\vc{W}$ than presented in this paper.}
\begin{subequations}
\begin{eqnarray}
\Pi_{xx}&=&-\eta_0^{\textrm{B}}(S_{xx}+S_{yy})\nonumber\\
& &-\eta_1^{\textrm{B}}(S_{xx}-S_{yy})-2\eta_3^{\textrm{B}}S_{xy}\label{Pxx}\\
\Pi_{yy}&=&-\eta_0^{\textrm{B}}(S_{xx}+S_{yy})\nonumber\\
& &+\eta_1^{\textrm{B}}(S_{xx}-S_{yy})+2\eta_3^{\textrm{B}}S_{xy}\\
\Pi_{xy}&=&-2\eta_1^{\textrm{B}}S_{xy}+\eta_3^{\textrm{B}}(S_{xx}-S_{yy})\\
\Pi_{xz}&=&-2\eta_2^{\textrm{B}}S_{xz}-2\eta_4^{\textrm{B}}S_{yz}\\
\Pi_{yz}&=&-2\eta_2^{\textrm{B}}S_{yz}+2\eta_4^{\textrm{B}}S_{xz}\\
\Pi_{zz}&=&-2\eta_0^{\textrm{B}}S_{zz}\label{Pzz},
\end{eqnarray}
\end{subequations}
where
\begin{subequations}
\begin{eqnarray}
\eta_0^{\textrm{B}}&=&  \mu_1       \label{Beta0}  \\
  \eta_1^{\textrm{B}} &=&   2\mu_2-\mu_1       \\
  \eta_2^{\textrm{B}} &=&   \mu_3       \\
  \eta_3^{\textrm{B}} &=&    \eta_1      \\
  \eta_4^{\textrm{B}} &=&     -\eta_2\label{Beta4}.      
\end{eqnarray}
\end{subequations}
Note that this simplification is possible only if $\zeta = 0$ and $\mu_v = 0$, which is not expected to be true in the strongly coupled ($\Gamma \gtrsim 1$) regime. 

One benefit of this organization is that it expresses the coefficients in terms of parallel, perpendicular, and cross components of the pressure tensor~\cite{1958JETP....6..358B}
\begin{equation}
\Pi=\Pi_{\parallel}+\Pi_{\perp}+\Pi_{\wedge}.
\end{equation}
This can be seen by writing Eqs.~(\ref{Pxx})-(\ref{Pzz}) in terms of products of $\vc{S}$ 
\begin{subequations}
\begin{eqnarray}
\label{PiPar}
\Pi_{\parallel}&=&-3\eta^{\textrm{B}}_0(\vc{b}\cdot\vc{S}\cdot\vc{b})\big(\vc{bb}-\frac{\vc{I}}{3}\big), \\
\label{PiP}
\Pi_{\perp}&=&-\eta^{\textrm{B}}_1W^{\prime}_{(1)}-\eta^{\textrm{B}}_2W^{\prime}_{(2)},\\
\label{PiW}
\Pi_{\wedge}&=&\frac{\eta^{\textrm{B}}_3}{2}W^{\prime\prime}_{(1)}+\eta^{\textrm{B}}_4W^{\prime\prime}_{(2)},
\end{eqnarray}
\end{subequations}
where $\vc{b}=\vc{B}/|\vc{B}|$ is the unit vector in the direction of the magnetic field, and
\begin{subequations}
\begin{eqnarray}
W^{\prime}_{(1)}&=&2(\vc{I}-\vc{bb})\cdot\vc{S}\cdot(\vc{I}-\vc{bb})\nonumber\\
& &-(\vc{I}-\vc{bb})(\vc{I}-\vc{bb}):\vc{S}\\
W^{\prime}_{(2)}&=&2(\vc{I}-\vc{bb})\cdot\vc{S}\cdot\vc{bb}+2\vc{bb}\cdot\vc{S}\cdot(\vc{I}-\vc{bb}) \\
W^{\prime\prime}_{(1)}&=&2\vc{b}\times\vc{S}\cdot(\vc{I}-\vc{bb})-2(\vc{I}-\vc{bb})\cdot\vc{S}\times\vc{b}\\
W^{\prime\prime}_{(2)}&=&2\vc{b}\times\vc{S}\cdot\vc{bb}-2\vc{bb}\cdot\vc{S}\times\vc{b}.
\end{eqnarray}
\end{subequations}
are traceless tensors. 
When organized in this way, the relations  $\vc{bb}:\Pi=\Pi_{\parallel}$, $\vc{bb}:\Pi_{\perp}=0$, $\vc{bb}:\Pi_{\wedge}=0$, and $\vc{S}:\Pi_{\wedge}=0$ show that Eq.~(\ref{PiPar}) involves velocity gradients parallel to $\vc{b}$,  Eq.~(\ref{PiP}) involves velocity gradients perpendicular to $\vc{b}$, and Eq.~(\ref{PiW}) involves velocity gradients perpendicular to both $\vc{b}$ and $\nabla \vc{V}$. 
Thus, the coefficient $\eta_0^{\textrm{B}}$ is related to the parallel stress, $\eta_1^{\textrm{B}}$ and $\eta_2^{\textrm{B}}$ to the perpendicular stress, and $\eta_3^{\textrm{B}}$ and $\eta_4^{\textrm{B}}$ to the cross stress.

\section{Molecular dynamics simulations\label{sec:MD}}

\subsection{Simulation Setup\label{sec:setup}}

Equilibrium MD simulations were carried out using the code LAMMPS~\cite{PLIMPTON19951}. In each simulation, the positions and velocities of 5000 particles were evolved in time through interaction via the Coulomb potential. The interaction was calculated using the particle-particle particle-mesh (P$^3$M) method~\cite{1984CoPhC..35..618E} with a short-range potential cutoff at $r=5a$. The particle-mesh calculation utilized a $75\times75\times75$ $k$-space mesh, with the mesh density chosen to ensure good energy conservation. Initialization at a chosen value of $\Gamma$ and $\beta$ involved fixing the number of particles, which scales the size of the periodic domain, followed by a 4000~$\omega_p^{-1}$ equilibration phase where particles achieved the desired temperature by using a Nos\'{e}-Hoover thermostat~\cite{1984JChPh..81..511N}. A time step of $\min \{0.01\omega_p^{-1},0.01\omega_p^{-1}/\beta\}$ was selected to ensure good energy conservation, resulting in a typical energy drift of $\lesssim 0.4\%$ over the $2\times 10^5 \omega_p^{-1}$ duration. The simulation was evolved in the NVT ensemble during which data was collected. 

Select simulations with greater particle number (up to 20,000), system size in the $z$-direction, and varying $k$-space mesh ($50^3$ to $75^3$) were used to ensure results were well-converged with respect to these parameters. Further convergence tests of the results are described in Sec.~\ref{sec:convergence}.

\subsection{Calculation of Viscosity~\label{sec:calc}} 

The viscosity coefficients $L_{\alpha\beta\gamma\delta}$ were calculated from correlations of fluctuations in the components of the stress tensor $\vc{\Pi}(t)$ by using the Green-Kubo relation~\cite{evans2007statistical} 
\begin{align}
&L_{\alpha\beta\gamma\delta}=\frac{1}{V k_B T}\times\nonumber\\
&\int_0^\infty dt \langle [\Pi_{\alpha\beta}(t)-\overline{PV}\delta_{\alpha\beta}][\Pi_{\gamma\delta}(0)-\overline{PV}\delta_{\gamma\delta}]\rangle.
\end{align}
Here, $\langle ... \rangle$ denotes an equilibrium ensemble average, 
and $\overline{PV}$ is the product of the pressure and system volume, which was computed from the long-time average of the diagonal elements of $\vc{\Pi}$~\cite{hansen2013theory}. 

The underlying physical process can be revealed by splitting the stress tensor into kinetic and potential components,
\begin{equation}\label{Pi1}
\Pi_{\alpha\beta}=\Pi^{\textrm{kin}}_{\alpha\beta}+\Pi^{\textrm{pot}}_{\alpha\beta},
\end{equation}
where
\begin{equation}\label{Pi2}
\Pi^{\textrm{kin}}_{\alpha\beta} \equiv  \frac{1}{V}\sum_{i=1}^Nm(\vc{v}_{i}\cdot\hat{\alpha}) (\vc{v}_{i}\cdot\hat{\beta})
\end{equation}
and
\begin{equation}\label{Pi3}
\Pi^{\textrm{pot}}_{\alpha\beta} \equiv \frac{1}{2V}\sum_{i=1}^N\sum_{j\ne i}^{N}\frac{(\vc{r}_{ij}\cdot\hat{\alpha})(\vc{r}_{ij}\cdot\hat{\beta})\phi^\prime(r_{ij})}{r_{ij}}.
\end{equation}
Here, $\hat{\alpha}$ is the unit vector in the $\alpha$ direction (e.g. $\hat{x}$), $\vc{r}_{ij}$ is the displacement vector from atom $i$ to atom $j$, and $\phi^\prime(r_{ij})$ is the derivative of the interaction potential with respect to $r_{ij}=|\vc{r}_{ij}|$. The separation of the pressure tensor in this way allows access to individual contributions associated with kinetic and potential components of transport coefficients. Such decompositions have proven useful in previous Green-Kubo-based calculations of shear viscosity and thermal conductivity of the OCP~\cite{OttPRE2015,DaligaultPRE2014}.  

One difficulty in calculating transport coefficients using the Green-Kubo formalism is the approximation of the ensemble average. Due to the finite nature of the simulation, the ensemble average is replaced with a finite time average of fluxes calculated over a finite spatial extent,
\begin{align}
&\langle [\Pi_{\alpha\beta}(t)-\overline{PV}\delta_{\alpha\beta}][\Pi_{\gamma\delta}(0)-\overline{PV}\delta_{\gamma\delta}]\rangle\approx C_{\alpha\beta\gamma\delta}(t,\tau)\nonumber\\
&\equiv\frac{1}{\tau}\int_0^\tau ds[\Pi_{\alpha\beta}(s)-\overline{PV}\delta_{\alpha\beta}][\Pi_{\gamma\delta}(s+t)-\overline{PV}\delta_{\gamma\delta}],\label{eq:corr}
\end{align}
where $\tau$ is the time series length and $C_{\alpha\beta\gamma\delta}(t,\tau)$ is a correlation function for a time series of duration $\tau$. The exact result is obtained in the limit where $\tau\to\infty$ and $\Pi_{\alpha\beta}(t)$ is calculated using an infinite system. For the purpose of calculating transport coefficients from simulation, it is sufficient to choose a system size that is large enough to avoid finite size effects and of long enough time duration $\tau$ for convergence of the correlation function. 
This will be discussed further in Sec.~\ref{sec:convergence}. 

\begin{figure*}
\includegraphics[width=\textwidth]{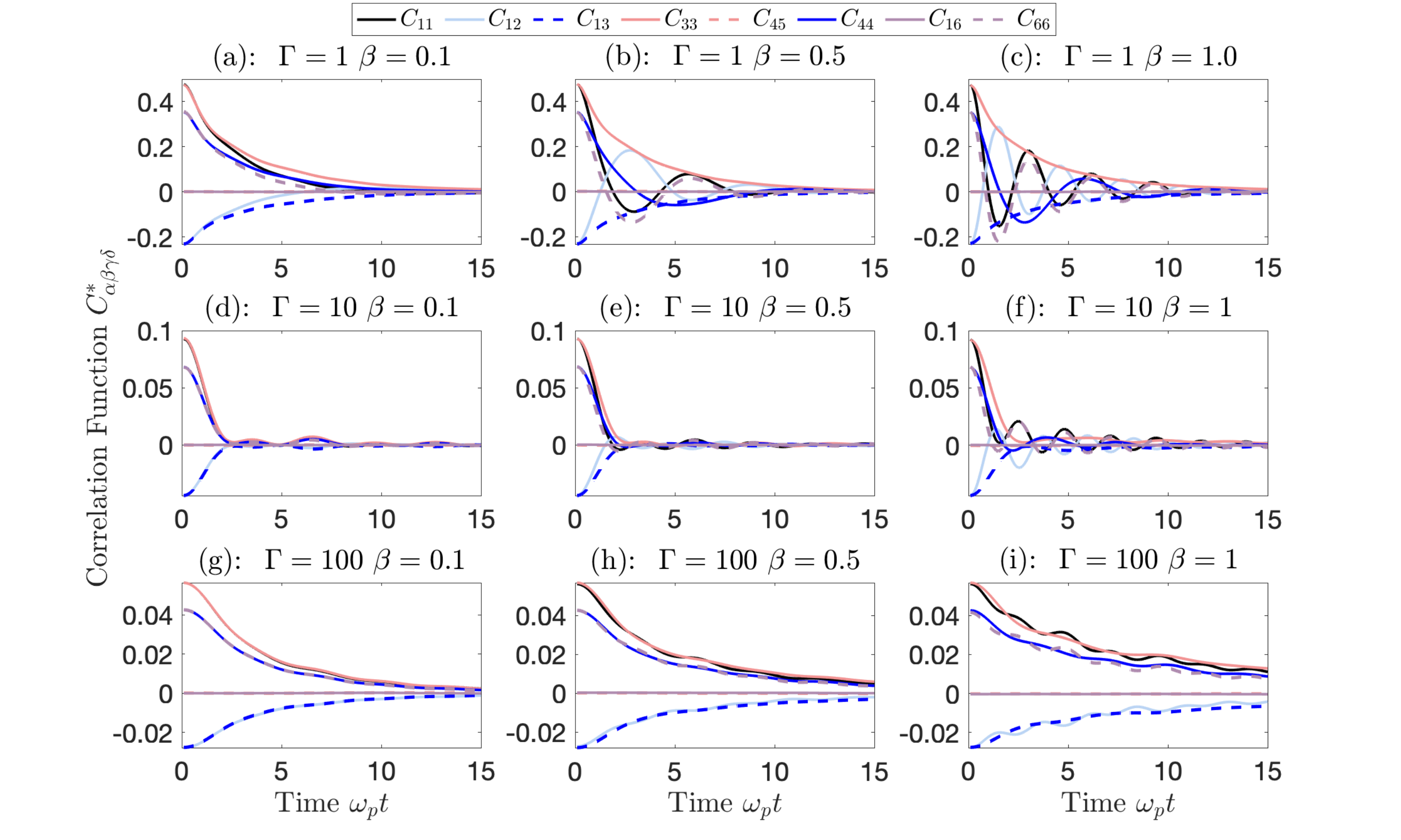}
\caption{\label{fig:acf} Correlation functions calculated from MD simulations for $\Gamma$=1, 10, and 100 with $\beta=$ 0.1, 0.5, and 1.0. The correlation functions shown are expected to be non-zero from the analysis of Sec.~\ref{sec:visc}.}
\end{figure*}

Since the magnetized OCP is characterized by two parameters, it is convenient to report the viscosity in the dimensionless units $L^*=L/mna^2\omega_p$ or $\eta^*=\eta/mna^2\omega_p$ so the values only depend on $\Gamma$ and $\beta$ implicitly. 
Similarly, the correlation functions are presented as $C_{\alpha\beta\gamma\delta}^*=C_{\alpha\beta\gamma\delta}/mna^2\omega_p$ which has units of $\omega_p$ so that their cumulative integral gives the corresponding dimensionless viscosity coefficient.

\subsection{Convergence of the Correlation Function\label{sec:convergence}}


The use of a correlation function of finite maximal time lag necessitates the truncation of the Green-Kubo integral at some time $t^*$. The value of $t^*$ is chosen so that the cumulative integral $\int_0^{t^*}dt C_{\alpha\beta\gamma\delta}(t,\tau)$ converges to a steady value. In practice, a large value of $t^*$ results in the integration of fluctuations at large time lags which are not physical, but statistical in nature, arising due to the finite nature of the stress tensor time series. Therefore $t^*$ is chosen to be the time needed for the correlation to decay to a near-zero value; in our simulations, $t^*\lesssim200\omega_p^{-1}$. See the correlation functions shown in Fig.~\ref{fig:acf} for characteristic decay times of each component.    

For times $t<t^*$, the lack of a converged correlation function can also corrupt the value of the cumulative integral. The convergence of the correlation function with $\tau$ can be split into two different components that can be checked separately: The convergence of the initial value with respect to the time lag $C_{\alpha\beta\gamma\delta}(t=0,\tau)$ and the convergence of the subsequent fluctuations at $t^*>t>0$. Convergence of each component must be satisfied to provide accurate values of transport coefficients.

The convergence of the $t=0$ value of $C_{\alpha\beta\gamma\delta}(0,\tau)$ with $\alpha=\gamma$, $\beta=\delta\ne\alpha$ can be verified by checking a series of relations following from the sum rule~\cite{DaligaultPRE2014} 
\begin{align}
 &\langle \Pi_{\alpha\beta}(0)\Pi_{\alpha\beta}(0)\rangle=N(k_BT)^2+\\
 &\frac{2\pi N n k_B T}{15 }\int_0^\infty dr r^3(g(r)-1)[4\phi^\prime(r)+r\phi^{\prime\prime}(r)]\nonumber
\end{align}
which has the following kinetic and potential components:
\begin{subequations}
\begin{eqnarray}
\label{sum1}
\langle \Pi^{\textrm{kin}}_{\alpha\beta}(0)\Pi^{\textrm{kin}}_{\alpha\beta}(0)\rangle &=& N(k_BT)^2\\
 \label{sum2}
 \langle\Pi^{\textrm{kin}}_{\alpha\beta}(0)\Pi^{\textrm{pot}}_{\alpha\beta}(0)\rangle &=& 0\\
\label{sum3}
\langle \Pi^{\textrm{pot}}_{\alpha\beta}(0) \Pi^{\textrm{pot}}_{\alpha\beta}(0)\rangle &=&\frac{2\pi N n k_B T}{15 }\int_0^\infty dr r^3(g(r)-1) \nonumber \\
 &\times & [4\phi^\prime(r)+r\phi^{\prime\prime}(r)].
\end{eqnarray}
\end{subequations}
These relations come from equilibrium statistical mechanics, and are independent of the magnitude or direction of $\vc{B}$. They hold for the Cartesian components $\alpha\beta = xy, xz$ and $yz$. 
Values for each of these components are shown in Table~\ref{table1} for $\Gamma=1$, $\beta=1$ and $\Gamma=10$, $\beta=1$. 
The calculated values indicate agreement with the sum rules to the fourth decimal place. 

\begin{table}
\centering
\begin{tabular}{lccc}
         $\alpha\beta \ \ \ $   & Eq.~(\ref{sum1})  & Eq.~(\ref{sum2}) & Eq.~(\ref{sum3})  \\ 
\hline\hline
 $\Gamma=1$ $\beta=1$   &  0.3333   &  0    &  0.0253      \\ \hline
    xy &      0.3292    &  $-1.40\times10^{-4}$     &  0.0253 \\
    xz &       0.3277    &  $-2.45\times10^{-4}$    &  0.0251   \\
    yz &       0.3278   &  $ \ 3.86\times10^{-5}$     &   0.0251  \\ \hline
$\Gamma=10$ $\beta=1$  &  0.0333    &  0  &   0.0354          \\\hline
    xy &      0.0329  &      $ \ 5.36\times 10^{-5}$    &  0.0353   \\
    xz &       0.0327  &    $ \ 6.95\times 10^{-6}$    &  0.0351   \\
    yz &       0.0327  &    $-1.50\times10^{-4}$   &  0.0354   \\ 
\end{tabular}
\caption{\label{table1} Comparison of the exact values of the right side of Eqs.~(\ref{sum1})-(\ref{sum3}) and those calculated from the corresponding correlation functions for the indicated $\alpha \beta$ components. }
\end{table}

\begin{figure}
\includegraphics[width=8cm]{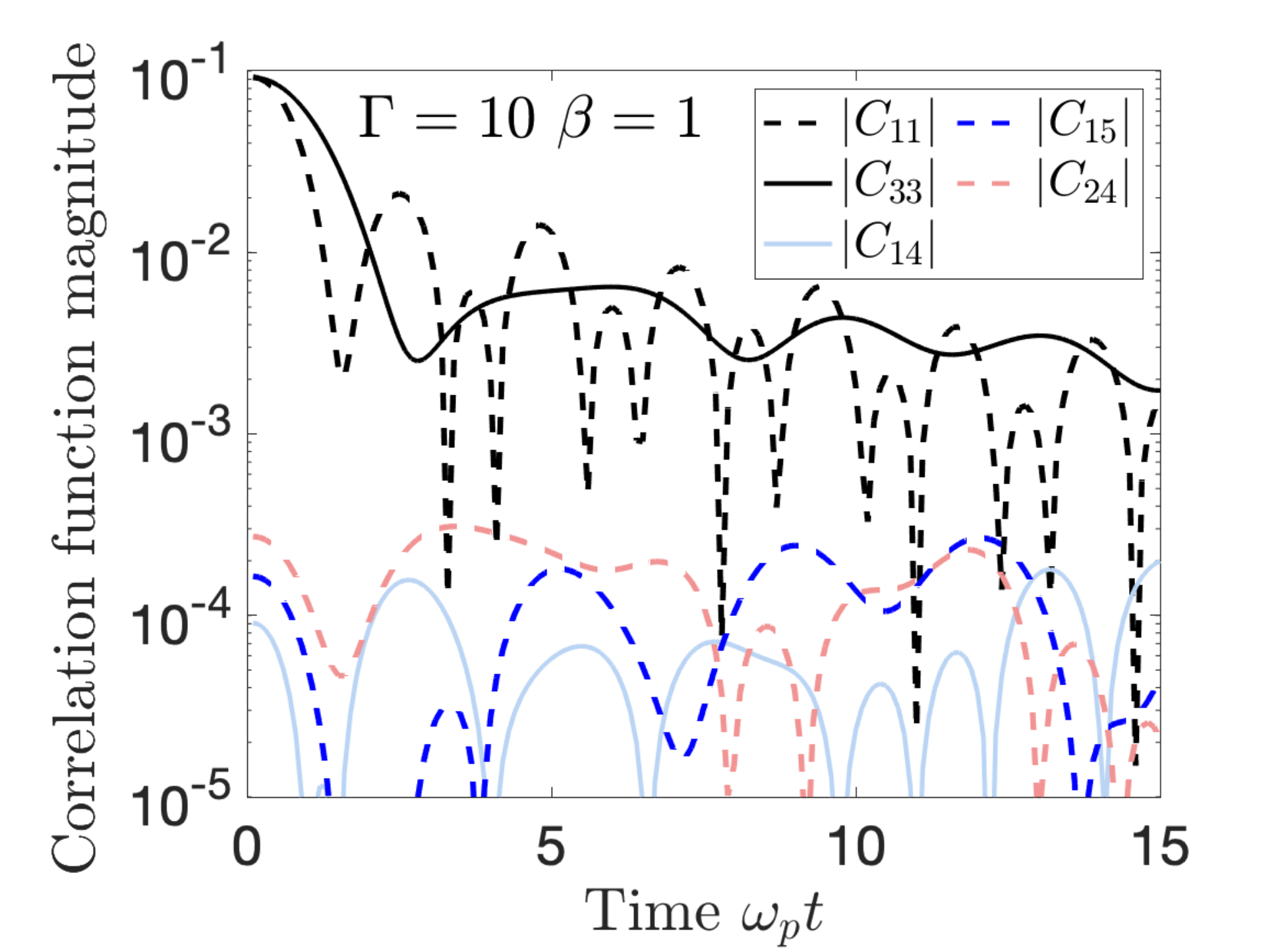}
\caption{\label{fig:error} A comparison of the magnitude of correlation functions $C_{14}$, $C_{15}$, and $C_{24}$, which are used as a measure of the error, with $C_{11}$ and $C_{33}$, which are expected to be non-zero in the thermodynamic limit. }
\end{figure}

Zwanzig and Ailawadi~\cite{1969PhRv..182..280Z} have provided an estimate of the error subsequent to the initial time that is associated with approximating an infinite time series by a finite one.
They estimate the second moment of the deviation from the exact value of the correlation as 
\begin{equation}\label{eqvar}
\la \Delta(t_1)\Delta(t_2) \ra \approx \frac{2\tau_e}{\tau}[C(0,\infty)]^2,
\end{equation}
where 
\begin{equation}
\Delta(t) \equiv C(t,\tau)-C(t,\infty).
\end{equation}
Here, $C(t,\infty)$ is the exact correlation function for an infinite time series, $\la ...\ra$ is the ensemble average, and $\tau_e$ is an estimate of the $1/e$ decay time of the correlation function. It follows that the statistical fluctuation level in the correlation function decreases as $1/\tau$. For the purposes of evaluating the noise level in the correlation function, it is useful to consider the variance of the fluctuations of components which are zero in the thermodynamic limit because these values deviate from zero significantly for time series of insufficient length and thus provides a good metric for the convergence of the correlation function with $\tau$~\cite{2012JChPh.137v4111H,PhysRevE.100.043206}. For the present calculation, several zero components of the correlation function are known from the symmetry arguments presented in Sec.~\ref{sec:visc}.  Three of these components, $C_{14}$, $C_{15}$, and $C_{24}$ are shown in Fig.~\ref{fig:error} alongside the non-zero components $C_{11}$ and $C_{33}$. The comparison demonstrates that the fluctuation level of the zero components is nearly two orders of magnitude lower than the non-zero components over the range shown. 
The numerical value of these coefficients is in the fourth decimal place, which is consistent with the numerical resolution resulting from the sum rule test shown in Table~\ref{table1}.
As a result of this low level of fluctuation the transport coefficients were not sensitive to $\sim20\%$ variations in the chosen value of $t^*$. 

\begin{figure}
\includegraphics[width=8.5cm]{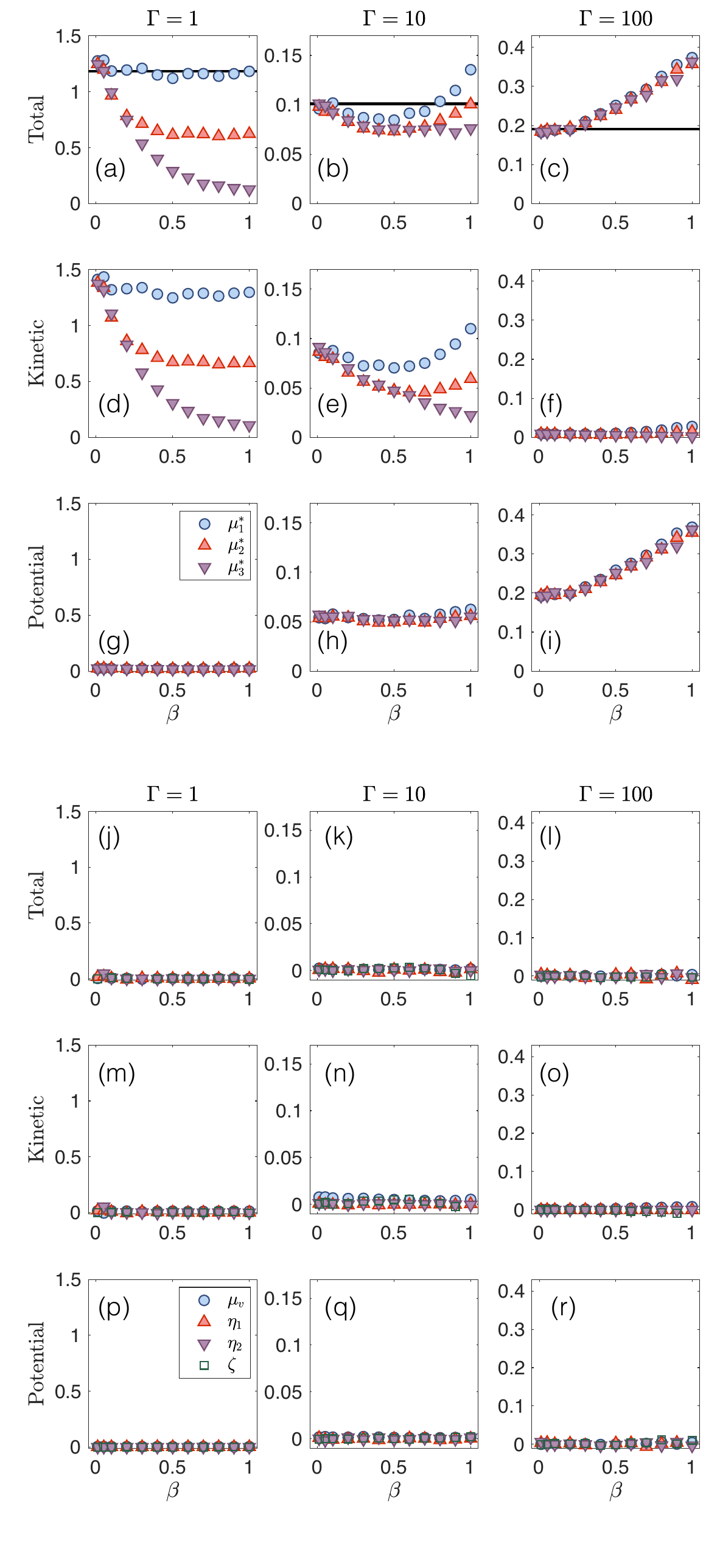}
\caption{\label{fig:CoefficientFig} Viscosity coefficients expressed in the form of Eqs.~(\ref{HDMmu1})-(\ref{HDMzeta}), along with the kinetic and potential components. Panels (a)-(i) show $\mu_1$, $\mu_2$ and $\mu_3$, and panels (j)-(r) show $\eta_1$, $\eta_2$, $\mu_v$ and $\zeta$. 
}
\end{figure}

\subsection{Results~\label{sec:results}}

Simulations were carried out for the conditions shown in Fig.~\ref{fg:gammabeta}, which includes $\Gamma$ ranging from 1 to 100 and $\beta$, from 0.01 to 2. 
Correlation functions for each component were calculated using Eq.~(\ref{eq:corr}) with Eqs.~(\ref{Pi1})-(\ref{Pi3}) used to split kinetic, potential and cross contributions; examples are shown in Fig.~\ref{fig:acf}. The values of the viscosity coefficients $L_{ij}$ were calculated as described in Sec.~\ref{sec:convergence} and are reported in Table ~\ref{tab:my-table}. From here, the coefficients can be put into either the form of Hooyman, DeGroot, and Mazur using Eqs.~(\ref{HDMmu1})-(\ref{HDMzeta}) or that of Braginskii using Eqs.~(\ref{Beta0})-(\ref{Beta4}). The coefficients are presented in Fig.~\ref{fig:CoefficientFig} for the former and Figs.~\ref{fig:braginskiicoeff} and \ref{fig:betascaling} for the latter of these forms.  
One should be cognizant that the Braginskii form is valid only if $\zeta = 0$ and $\mu_v =0$, but this will be shown to be consistent with our simulation results to within the attained numerical accuracy. 

\begin{table*}
\caption{Computed values of the viscosity coefficients.}
\label{tab:my-table}
\begin{tabular}{lllllllll}
 & $L_{11}^*$ & $L_{12}^*$ & $L_{13}^*$ & $L_{33}^*$ & $L_{44}^*$ & $L_{45}^*$ & $L_{16}^*$ & $L_{66}^*$ \\ \hline\hline
 $\Gamma=1$&     &     &     &     &     &     &     &     \\ \hline
$\beta=0.01$ &    1.6479  & -0.7877 &  -0.8472 &   1.7005  &  1.2446 &  -0.0244&    0.0161 &   1.2543   \\
$\beta=0.05$ &    1.5543 &  -0.6605&   -0.8827 &   1.6322  &  1.1843&    0.0486&    0.0215 &   1.1285    \\
 $\beta=0.1$&     1.1406   &-0.3485  & -0.7805   &  1.5965   &0.9940 &   0.0096 &   0.0050 &   0.7750    \\
 $\beta=0.2$&     0.7672  &  0.0298 &  -0.7901  &  1.6005   & 0.7479 &   0.0015&   -0.0072 &   0.3850      \\
 $\beta=0.3$&    0.6286   & 0.1946  & -0.8134    &1.5853    & 0.5354  &  0.0009&    0.0060  &   0.2126    \\
 $\beta=0.4$&     0.5317   & 0.2383 &  -0.7642   & 1.5350  &  0.3991 &  -0.0006 &   0.0042 &   0.1452     \\
$\beta=0.5$ &   0.4816    &0.2660   &-0.7414     &1.4999    &0.2906   & 0.0023    &0.0014   &  0.1103\\
$\beta=0.6$ &    0.4787   & 0.2958  & -0.7733    &1.5498   & 0.2331   & 0.0000   &-0.0010  &  0.0776    \\
$\beta=0.7$ &    0.4692   & 0.3100   &-0.7731    &1.5464    &0.1768    &0.0002    &0.0002    &0.0711   \\
$\beta=0.8$ &    0.4430   & 0.3133   &-0.7521    &1.5311    &0.1612    &0.0046    &0.0013    &0.0708   \\
$\beta=0.9$ &     0.4490   & 0.3230   &-0.7659   & 1.5613   &0.1382    &0.0005   &-0.0020   & 0.0665  \\
$\beta=1.0$ &     0.4635    &0.3356   &-0.7860   & 1.5725   & 0.1268    &0.0005    &0.0020   & 0.0652   \\
$\beta=1.5$&     0.5856    &0.4627   &-1.0035    &1.9996    &0.0873    &0.0003   &-0.0008   & 0.0691\\
$\beta=2.0$&    0.7829    &0.6273   &-1.3129    &2.5079    &0.0665   &-0.0004   &-0.0031   & 0.0716\\ \hline
 $\Gamma=10$&     &     &     &     &     &     &     &     \\ \hline
$\beta=0.01$ &    0.1338  & -0.0667   &-0.0611    &0.1318    &0.1013   & 0.0008   & 0.0006   & 0.0992     \\
$\beta=0.05$ &   0.1221   &-0.0546   &-0.0631    &0.1321     &0.0992  &  0.0014   & 0.0017   & 0.1000     \\
 $\beta=0.1$&      0.1194   &-0.0484   &-0.0662    &0.1375   & 0.0922  &  0.0001   & 0.0018   & 0.0862     \\
 $\beta=0.2$&     0.1061   &-0.0423   &-0.0597    &0.1224   & 0.0842   & 0.0015    &0.0014    & 0.0726     \\
 $\beta=0.3$&     0.0942   &-0.0362   &-0.0560    &0.1199   & 0.0787   & 0.0005   &-0.0005  &  0.0702     \\
 $\beta=0.4$&     0.0912   &-0.0343   &-0.0551    &0.1181   & 0.0751      &0.0015   &-0.0023   & 0.0616    \\
$\beta=0.5$ &     0.0894   &-0.0341   &-0.0539    &0.1180    &0.0759     & 0.0007 &   0.0011   & 0.0621    \\
$\beta=0.6$ &     0.0900   &-0.0309   &-0.0585    &0.1277    &0.0746   & 0.0000   & 0.0000   & 0.0619   \\
$\beta=0.7$ &     0.0932   &-0.0317   &-0.0601    &0.1291     &0.0750    & 0.0020  &  0.0009   & 0.0636    \\
$\beta=0.8$ &     0.0999   &-0.0297   &-0.0675    &0.1402    &0.0761    &0.0022   & -0.0016   & 0.0628    \\
$\beta=0.9$ &     0.1081   &-0.0268   &-0.0767    &0.1495     &0.0721   &-0.0003   & -0.0022  &  0.0651    \\
$\beta=1.0$ &     0.1161   &-0.0148     &-0.0905    &0.1750    &0.0761   & 0.0003    & 0.0014    & 0.0706    \\\hline
 $\Gamma=100$&     &     &     &     &     &     &     &     \\ \hline
$\beta=0.01$ &    0.2479   &-0.1236   &-0.1231    &0.2405   & 0.1834   & 0.0031    &0.0044    &0.1878    \\
$\beta=0.05$ &   0.2512   &-0.1266   &-0.1234    &0.2549   & 0.1840   &-0.0018    &0.0024    &0.1848     \\
 $\beta=0.1$&     0.2500   &-0.1248   &-0.1240    &0.2537   & 0.1910    &-0.0008   &-0.0004   & 0.1883    \\
 $\beta=0.2$&     0.2614   &-0.1347   &-0.1260    &0.2556   & 0.1903   & 0.0025    &0.0020    &0.1933     \\
 $\beta=0.3$&     0.2715   &-0.1313   &-0.1385     &0.2831   & 0.2045   &-0.0012   &-0.0046    &0.2033     \\
 $\beta=0.4$&     0.2973   &-0.1387  & -0.1551     &0.3001   & 0.2289   &-0.0033   &-0.0111    &0.2104    \\
$\beta=0.5$ &     0.3134   &-0.1434   &-0.1680     &0.3328   & 0.2464   &-0.0015    &0.0037   & 0.2355    \\
$\beta=0.6$ &     0.3545   &-0.1654  & -0.1812     & 0.3632   & 0.2669   &-0.0005   & 0.0046   & 0.2553    \\
$\beta=0.7$ &     0.3873   &-0.1857   &-0.1939    &0.3890   & 0.2781     &0.0053   &-0.0083    &0.2807    \\
$\beta=0.8$ &     0.4071   &-0.1866   &-0.2119     &0.4426    & 0.3156   &-0.0031    &0.0016    &0.3153    \\
$\beta=0.9$ &     0.4583   &-0.2030   &-0.2402   & 0.4583   & 0.3189    &0.0071    &0.0073    &0.3287    \\
$\beta=1.0$ &      0.4714   &-0.2102   &-0.2465    &0.4976    & 0.3631    &-0.0034   &-0.0100   &0.3341   \\
$\beta=1.5$& 0.6103       & -0.2667   &-0.3143    &0.6149    &0.4419    &0.0041    &0.0011    &0.4462\\
$\beta=2.0$&0.6759      &-0.3223   &-0.3406    &0.6769    &0.4951    &0.0051    &0.0019    &0.4924\\
\end{tabular}
\end{table*}

\section{Discussion\label{sec:discussion}}

Results demonstrate that qualitative changes in the viscosity coefficients depend on both $\beta$ and $\Gamma$. 
This section discusses significant trends. 
Section~\ref{sec:acf} discusses how the differences in each viscosity coefficient stem from anisotropies of the fluctuations, as described by the correlation functions, that arise when the magnetic field is sufficiently strong. 
Section~\ref{sec:uc} discusses how the numerical resolution of the simulations was sufficient to resolve the shear viscosity coefficients associated with parallel and perpendicular shear stresses, but was unable to resolve the shear viscosity coefficients associated with cross component of the shear stress, or the bulk or cross coefficients. 
Section~\ref{sec:rt} shows that changes in scaling of the shear viscosity coefficients with $\beta$ occur at the regime transitions predicted in Fig.~\ref{fg:gammabeta}, and that results at $\Gamma = 1$ are consistent with the scaling predicted by the Braginskii equations over the narrow range of the classical magnetized regime that was accessed by the simulations.  
Section~\ref{sec:gm10} compares the kinetic and potential contributions to the viscosity coefficients, showing that the transition between the dominance of one component over the other depends on $\beta$ as well as $\Gamma$. 
Finally, Sec.~\ref{sec:gm100} shows that at strong coupling ($\Gamma = 100$) all three of the resolved shear viscosity coefficients merge to a common value, regardless of the magnetization strength. 


\subsection{Anisotropy of fluctuations\label{sec:acf}}

The correlation functions $C_{\alpha\beta\gamma\delta}$ demonstrate the most basics properties of anisotropy of the fluid stress fluctuations resulting from the application of an external magnetic field; as shown in Fig.~\ref{fig:acf}. 
In the unmagnetized limit, it is expected that the fluctuations in stress are independent of choice of coordinate axis. 
For example, one expects that $C_{11}=C_{33}$ since these correspond to the autocorrelation of stress fluctuations of the $xx$ or $zz$ components.  Likewise, it is also expected that $C_{12}=C_{13}$ due to symmetry between $yy$ and $zz$ and $C_{44}=C_{66}$ due to symmetry between $yz$ and $xy$. These relations are easily verified in the unmagnetized cases shown in Figs.~\ref{fig:acf}(a), (d), and (g), aside from slight deviations due to the weak magnetic field at $\beta=0.1$. These symmetry relations are independent of the value of $\Gamma$.

Considering $\Gamma=1$, the anisotropy becomes apparent as $\beta$ increases to 0.5. Figures ~\ref{fig:acf}(b) and (c) show that correlations between tensor components with at least one coordinate (index) in the plane perpendicular to the magnetic field ($x$ or $y$ directions) exhibit oscillations. 
These are associated with gyromotion, and the oscillation frequency is characterized by $\omega_c$. 
For example, as the magnetization doubles from $\beta=0.5$ in Fig.~\ref{fig:acf}(b) to $\beta=1.0$ in Fig.~\ref{fig:acf}(c) the period of the oscillation also doubles. 
It is also noteworthy that the gyromotion causes these components to oscillate between positive and negative correlations, whereas they are of a definite sign in an unmagnetized weakly coupled plasma. Because the viscosity coefficients are the time integrals of the correlation functions, the result of the oscillations is a significant reduction of the resulting coefficients, as shown in Table~\ref{tab:my-table}.

Considering $\Gamma=10$, the oscillations associated with gyromotion are strongly suppressed in comparison to $\Gamma=1$. 
The reason for this is that the gyromotion influences the kinetic components of the stress significantly, but not the potential components. 
The ratio of the kinetic component to the potential component is much smaller at $\Gamma = 10$ than at $\Gamma = 1$. 

\begin{figure}
\includegraphics[width=\columnwidth]{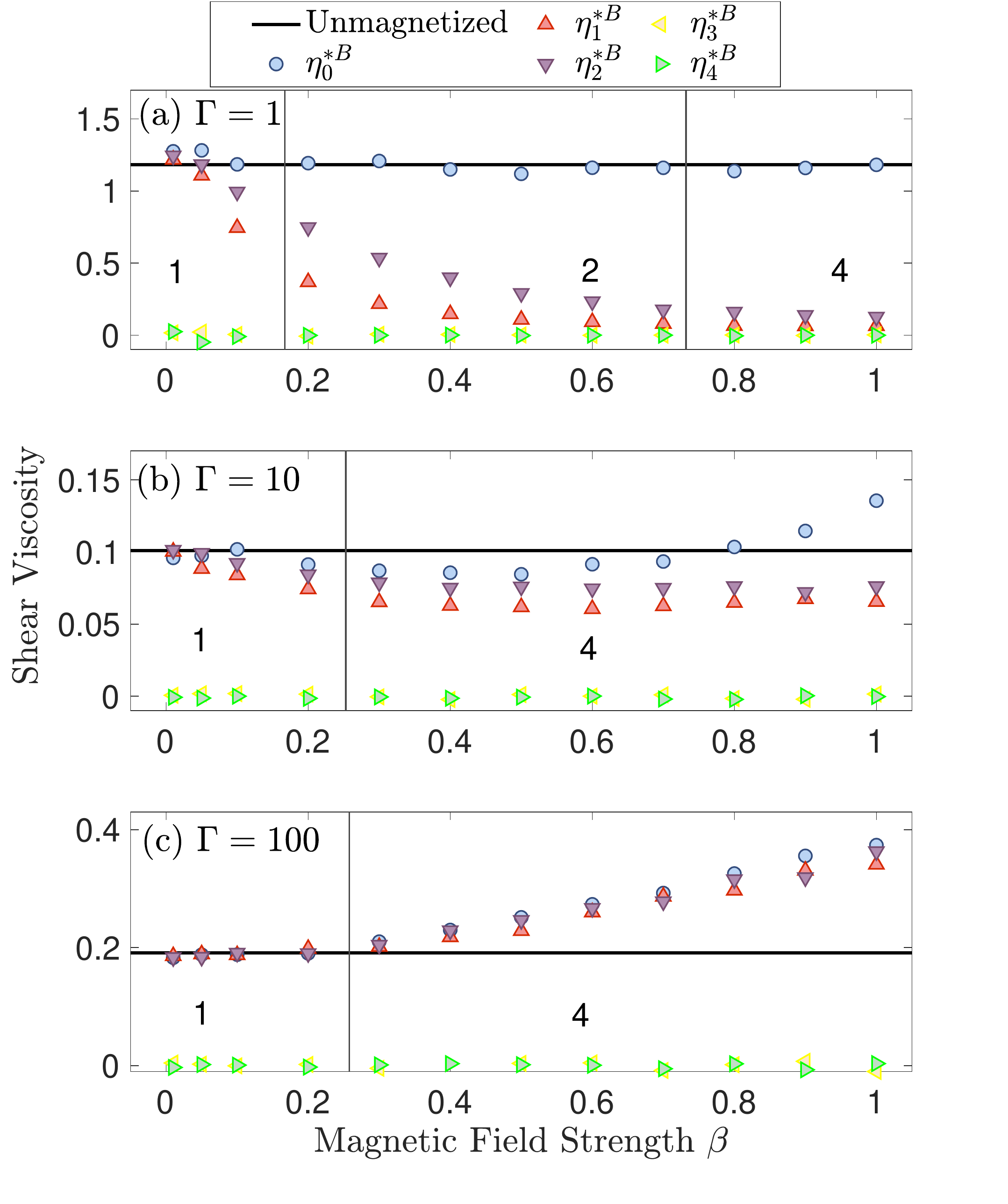}
\caption{\label{fig:braginskiicoeff}Dependence of the five shear viscosity coefficients on the magnetization parameter for three values of the Coulomb coupling strength: (a) $\Gamma=1$, (b) $\Gamma =10$ and (c) $\Gamma =100$. The vertical lines denote the boundaries in Fig.~\ref{fg:gammabeta}, and horizonal lines the $\beta=0$ results from~\cite{DaligaultPRE2014}.}
\end{figure}

As $\Gamma$ increases to 100, the kinetic component of the stress becomes insignificant. The correlation for $\Gamma=100$ exhibit the same symmetry properties as the unmagnetized case, with the exception of a very slight oscillation at $\beta=1$ (cf. Fig.~\ref{fig:CoefficientFig}(f) which shows a small non-zero contribution to the kinetic portion of the transport coefficient at $\beta=1$). However, the overall magnitude of the correlation function tail increases with $\beta$. This suggests that the magnetization acts to increase the transport rates, but does not increase the anisotropy of the fluctuations of fluid stress.

\subsection{Unresolved coefficients\label{sec:uc}} 

Figure~\ref{fig:CoefficientFig} shows the viscosity coefficients expressed in the form of Eqs.~(\ref{HDMmu1})-(\ref{HDMzeta}). 
Panels (j)-(l) show that both of the shear viscosity coefficients associated with the cross component of the shear stress ($\eta_1$ and $\eta_2$), as well as the bulk viscosity ($\mu_v$) and cross coefficient ($\zeta$) are consistent with zero. 
Although each of these is expected to be smaller than the shear viscosity coefficients shown in panels (a)-(c), none of these coefficients are expected to be identically zero. 
The result is likely due to the achievable numerical resolution of the MD simulations, as discussed in Sec.~\ref{sec:convergence}. 

For example, Braginskii transport predicts that the shear viscosity coefficients associated with the cross component of the shear stress are negligible in the unmagnetized regime, scaling as $\omega_c/\nu_\textrm{col}$ for $\omega_c/\nu_\textrm{col} \ll 1$ where $\nu_{\textrm{col}}$ is the Coulomb collision frequency, and also decrease with magnetic field strength in the classically magnetized regime as $(\omega_c/\nu_{\textrm{coll}})^{-1}$ for $\omega_c/\nu_\textrm{col} \gg 1$. 
These coefficients peak at $\omega_c/\nu_{\textrm{coll}} \approx 1$, but even then are expected to take values that are smaller than the other shear viscosity coefficients. 
Although these coefficients are not expected to be zero, they are not resolved by the MD computations. 
In addition to the coefficients themselves, Fig.~\ref{fig:acf} shows that the correlation functions from which these coefficients are computed ($C_{16}$ and $C_{45}$) are nearly zero at all times.

\begin{figure}
\includegraphics[width=\columnwidth]{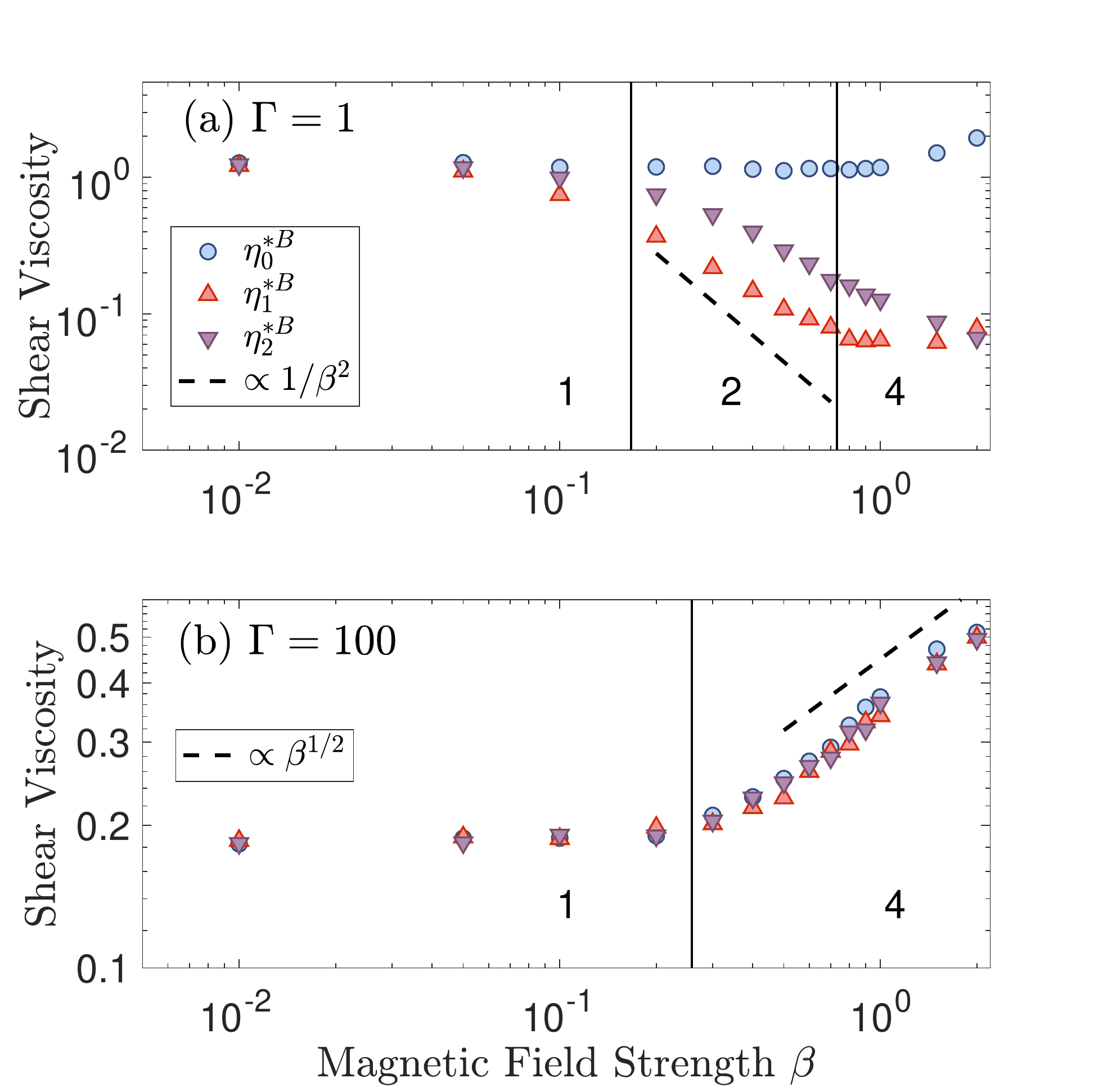}
\caption{\label{fig:betascaling}Shear viscosity coefficients $\eta_o^{*B}$, $\eta_1^{*B}$ and $\eta_2^{*B}$ computed from MD simulations for $\Gamma=1$ (top) and $\Gamma = 100$ (bottom) as the magnetization parameter $\beta$ varies through the unmagnetized regime (1), the classically magnetized regime (2) and the extremely magnetized regime (4).}
\end{figure}

\begin{figure}
\includegraphics[width=\columnwidth]{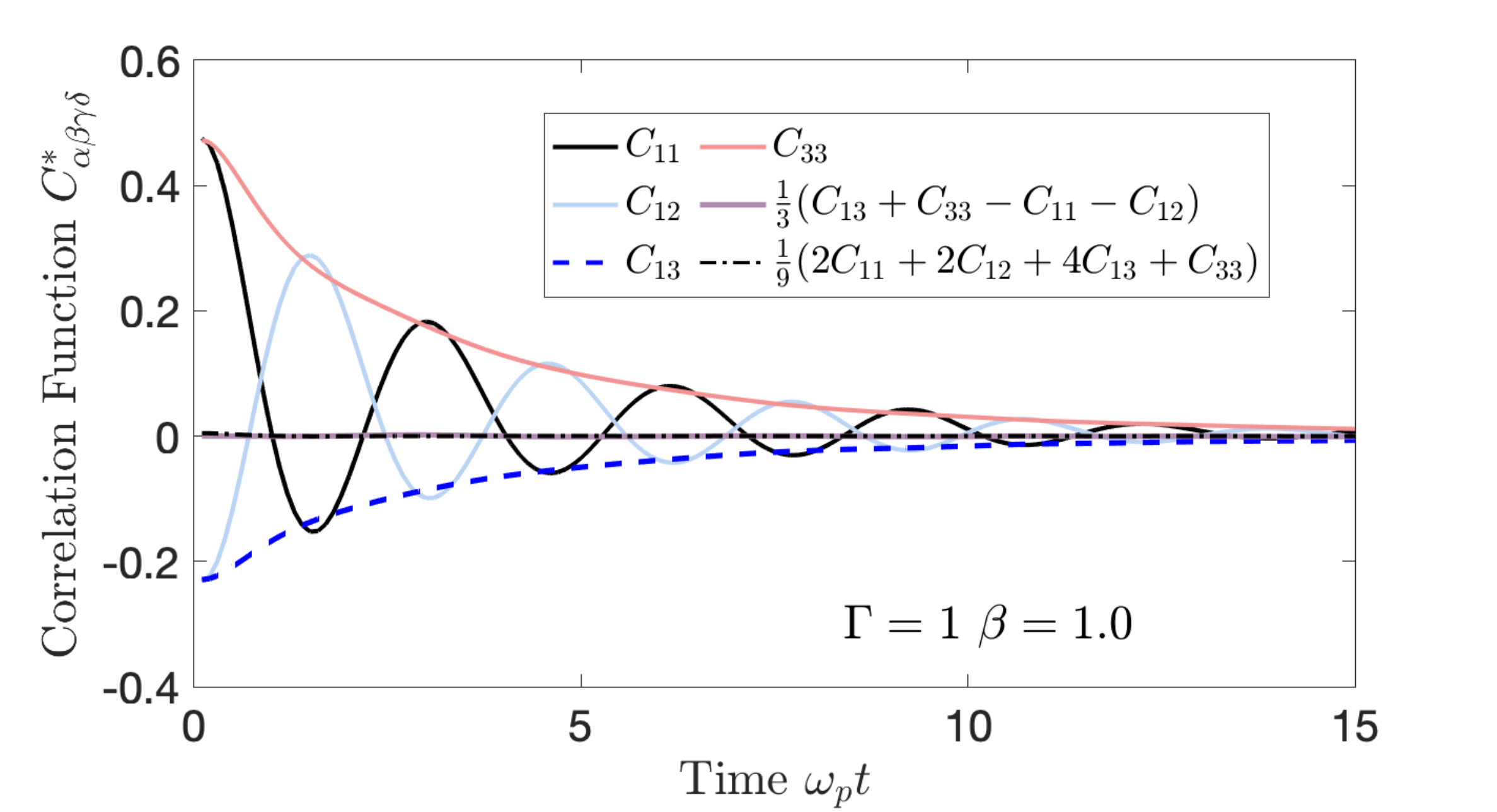}
\caption{\label{fig:bulk}The individual and sum of the correlation functions whose cumulative integral corresponds to the bulk viscosity in Eq.~(\ref{HDMBulk}) and cross viscosity in Eq.~(\ref{HDMzeta}).}
\end{figure}

Similarly, bulk viscosity is not expected to be zero in general. 
Traditional weakly coupled plasma theory resulting from the Chapman-Enskog solution of the plasma kinetic equation predicts that both the bulk and cross terms are identically zero~\cite{chapman1939mathematical}. 
However, this is a consequence of the weakly coupled limit assumed in the kinetic theory, and is not expected to be a general result. 
Nevertheless, early MD simulations of the unmagnetized OCP by Vieillefosse and Hansen~\cite{PhysRevA.12.1106} showed that the bulk viscosity coefficient is also negligible compared to shear viscosity for $\Gamma$ values ranging from 1 to 160. 
It was unknown how magnetization should influence this result. 
However, unlike the cross field components which depend on $C_{16}$ and $C_{45}$, the bulk viscosity ($\mu_v$) and cross coefficient ($\zeta$) are near-zero due to a precise cancellation of non-zero terms. Figure~\ref{fig:bulk} shows how the sum of non-zero correlation functions, corresponding to Eq.~(\ref{HDMBulk}) and Eq.~(\ref{HDMzeta}), sum to a correlation function that is nearly zero, the cumulative integral of which is also near-zero. 

\subsection{Regime transitions\label{sec:rt}} 

Fundamental transitions in the scaling of transport coefficients with $\beta$ have been predicted to occur at the boundaries indicated in Fig.~\ref{fg:gammabeta}~\cite{BaalrudPRE2017}, which are defined by comparing the gyroradius with other relevant physical scales in the system, as described in the introduction. 
They have been previously tested by comparing with MD simulations of diffusion and temperature anisotropy relaxation rates~\cite{BaalrudPRE2017}. 
Figures~\ref{fig:braginskiicoeff} and \ref{fig:betascaling} show that these boundaries also predict where the transitions in the shear viscosity coefficients occur. 

Focusing on the logarithmic scale in Fig.~\ref{fig:betascaling}, which shows the three coefficients $\eta_o^\textrm{B}$, $\eta_1^\textrm{B}$ and $\eta_2^\textrm{B}$ in the Braginskii form from Eqs.~(\ref{Beta0})-(\ref{Beta4}), all three coefficients merge to the same value, independent of $\beta$ in the unmagnetized regime (region 1). 
In this limit, the shear viscosity tensor can be reduced to a single scalar coefficient, as expected from the symmetry of an unmagnetized plasma. 
The coefficients obtained in this limit agree well with the previous results from~\cite{DaligaultPRE2014}. 

Considering $\Gamma = 1$, as $\beta$ increases into the classically magnetized regime (region 2), the coefficient associated with parallel stress $\eta_o^\textrm{B}$ remains unchanged, while the two coefficients associated with perpendicular stress $\eta_1^\textrm{B}$ and $\eta_2^\textrm{B}$ both decrease sharply with increasing $\beta$. 
Recall that this reduction is associated with oscillations in the corresponding correlations functions, as shown in Fig.~\ref{fig:acf}. 
This is the classically magnetized regime (region 2), in which the Braginskii scaling arguments are expected to hold: $\eta_0^\textrm{B}\propto\beta^0$, $\eta_1^\textrm{B}\propto\beta^{-2}$, $\eta_2^\textrm{B}\propto\beta^{-2}$.
The data shown in Fig.~\ref{fig:betascaling} appear to be consistent with these predictions.  
The scaling of $\eta_1^\textrm{B}$ and $\eta_2^\textrm{B}$ are somewhat more gradual than $\beta^{-2}$, but the range of $\beta$ values corresponding to region 2 is narrow (less than one decade) at $\Gamma = 1$. 
The more gradual scaling that is observed is likely due to a transition to region 4, where a flattening of the scaling with $\beta$ is observed. 
A more rigorous test of the Braginskii formulas would require simulations at a much lower $\Gamma$ value, but these are much more computationally expensive. 

As the $\beta$ value increases into region 4 for $\Gamma=1$, the $\eta_o^\textrm{B}$ coefficient becomes dependent on $\beta$, increasing as a positive power. 
The coefficient $\eta_1^\textrm{B}$ flattens dramatically, becoming nearly independent of $\beta$, or perhaps scaling with a slightly positive exponent of $\beta$. 
The third shear viscosity coefficient $\eta_2^\textrm{B}$ scales somewhat more gradually with $\beta$ than in region 2, but more more steeply than does the $\eta_1^\textrm{B}$ coefficient. 
There is currently no satisfactory kinetic theory to describe region 4, so these MD results provide a unique first-principles computation that future theoretical developments can use as a benchmark. 

At $\Gamma = 100$, only two regions (1 and 4) are predicted. 
The data shown in Fig.~\ref{fig:betascaling} is consistent with this, showing that all shear viscosity coefficients are independent of $\beta$ in region 1, and transition to a positive scaling with $\beta$ (approximately as $\beta^{1/2}$) in region 4. 
A similar result has been noted for the parallel component of thermal conductivity in the strongly magnetized Yukawa-screened OCP~\cite{OttPRE2015}. This effect was attributed to the existence of more frequent collisions in the presence of a strong magnetic field. When a strong field is present, particles move along their field line and collide with larger collision angles due to this magnetization. A similar process may increase the field parallel viscosity.

\subsection{Potential and kinetic contributions\label{sec:gm10}}

An advantage of computing transport coefficients using the Green-Kubo relations is that they reveal the relative contributions from particle momenta (kinetic contributions) and direct interactions (potential contributions). 
Figure \ref{fig:CoefficientFig} shows a breakdown of each component for the viscosity coefficients. 
As has been observed for the unmagnetized case~\cite{DaligaultPRE2014}, at $\Gamma=1$ the shear viscosity is entirely due to the kinetic component. 
Since the particle momenta are significantly influenced by magnetization, causing oscillations in associated components the correlation functions as shown in Fig.~\ref{fig:acf}, magnetization significantly reduces the kinetic components of $\mu_2$ and $\mu_3$. 
Because it is based on a Boltzmann kinetic equation, the Braginskii theory only accounts for the kinetic components of transport coefficients, which is an accurate approximation at weak coupling.



Figures~\ref{fig:CoefficientFig}(e) and (h) show that at $\Gamma = 1$, both kinetic and potential components contribute to the total viscosity. 
The same observation has been made in the unmagnetized case~\cite{DaligaultPRE2014}, where it was shown that the transition point between dominance of kinetic and potential components occurs at the minimum of the viscosity coefficient at $\Gamma \approx 17$. 
Here, it is observed that both $\beta$ and $\Gamma$ influence the viscosity coefficients, and that it influences each in a quantitatively different way. 
Magnetization causes non-monotonic changes to the kinetic components of $\mu_1$, $\mu_2$, and $\mu_3$. 
It is also interesting to notice that the potential components of $\mu_1$, $\mu_2$, and $\mu_3$ are nearly equal, regardless of the $\Gamma$ or $\beta$ values. 

Finally, at the strongest coupling condition of $\Gamma = 100$, the shear viscosity is entirely determined by the potential contributions; the kinetic contributions being negligible. 
The potential contribution of each coefficient is observed to increase as a positive power of $\beta$ (approximately $\beta^{1/2}$ over this range).

\subsection{Merging of coefficients at strong coupling\label{sec:gm100}}

The most striking feature of the shear viscosity coefficients at $\Gamma = 100$ is that they merge to a common value $\mu_1=\mu_2=\mu_3$; see Fig.~\ref{fig:CoefficientFig}(c) and Fig.~\ref{fig:betascaling}(b). 
As Fig.~\ref{fig:betascaling} shows, the potential components of $\mu_1$, $\mu_2$ and $\mu_3$ are the same at \emph{all} values of $\Gamma$ and $\beta$ simulated. 
When $\Gamma$ is sufficiently large, the potential components are much larger than the kinetic components and so determine the total shear viscosity. 
Thus, it is reasonable to expect that the merging of coefficients ($\mu_1=\mu_2=\mu_3$) is associated with the predominance of the potential contributions, which is a strong coupling effect. 

The equality of the shear viscosity coefficients stems from the near equality of the relevant components of the correlation functions shown in Fig.~\ref{fig:acf}(g)-(i), $C_{12}\approx C_{13}$ and $C_{11}\approx C_{33}$, as discussed in Sec.~\ref{sec:acf}. With these relations, Eqs.~(\ref{HDMmu1}) and (\ref{HDMmu2}) result in $\mu_1=\mu_2$. Since the correlation functions exhibit the same symmetries expected of stress fluctuations in an isotropic system, it is expected that $\mu_1=\mu_3$ as well.



\section{Conclusion\label{sec:conclusion}}

This paper presented calculations of the coefficients of the viscosity tensor in a magnetized strongly coupled plasmas using equilibrium molecular dynamics simulations. The results were analyzed in three different magnetization regimes set by length scales in the plasma: (1) the unmagnetized regime where $\lambda_\textrm{col}<r_c$, (2) the classically magnetized regime where $r_c<\lambda_\textrm{col}$ and $r_c$ is still greater than $\lambda_D$ and $r_L$, and (3) where $r_c$ is the smallest length scale in the plasma. Qualitative differences in the shear viscosity coefficients were observed in each of these regimes in agreement with Ref.~\cite{BaalrudPRE2017}. 

In the unmagnetized regime, the shear viscosity tensor reduces to a single scalar coefficient, consistent with expectations due to symmetries in this limit. In the classical magnetized regime, the shear viscosity coefficient associated with the stress in the field parallel direction are unmodified, while those associated with the perpendicular stress decrease with increasing beta in a manner consistent with the predictions of Braginskii transport. 
In the extremely magnetized regime, the viscosity coefficients are observed to increase, rather than decrease, with beta. At large values of $\Gamma$ the transport coefficients associated with parallel and perpendicular stress are observed to merge to a single coefficient. In this case, inspection of the correlation functions indicate that the plasma does not exhibit anisotropic fluctuations in the fluid stress. 

The data provided here may be useful as a benchmark for the evaluation of strongly coupled plasma theories in each of these three magnetization regimes. While the data is constraining for Braginskii theory, the narrowness of region 2 and the influence of the transition between regimes at $\Gamma=1$ prevents exact confirmation of Braginskii theory from the MD data. Future simulations at weaker coupling may better probe this regime.  
\section*{ACKNOWLEDGEMENTS}
This work was supported by the U.S. Department of Energy, Office of Fusion Energy Sciences, under Award No. DE-SC0016159.

\section*{Appendix: Coordinate Rotations of Symmetric Rank 4 Tensors}

While the transformation rules of symmetric rank 4 tensors using Voigt notation are well known in some areas such as the design of piezoelectric materials~\cite{bao2005analysis}, they are uncommon in plasma physics. This section presents a quick review aimed towards the problem presented in this paper. A more complete discussion can be found in Chapter 6 of Ref.~\cite{bao2005analysis}. 

First consider the transformation of a rank 2 Cartesian tensor under a general coordinate rotation
\begin{equation}\label{eq:matrixtransform}
A^\prime=R A R^{-1},
\end{equation}
where
\begin{equation}
A=
\begin{pmatrix}
A_{xx} & A_{xy} & A_{xz}\\
A_{yx} & A_{yy} & A_{yz}\\
A_{zx} & A_{zy} & A_{zz}
\end{pmatrix}
\end{equation}
and
\begin{widetext}
\begin{align}
R=&
\begin{pmatrix}
l_1 & m_1 & n_1\\
l_2 & m_2 & n_2\\
l_3 & m_3 & n_3
\end{pmatrix}=
&\begin{pmatrix}
\cos\psi\cos\theta\cos\phi-\sin\psi\sin\phi \ \ & \cos\psi\cos\theta\sin\phi+\sin\psi\cos\phi \ \  & -\cos\psi\sin\theta\\
-\sin\psi\cos\theta\cos\phi-\cos\psi\sin\phi \ \ & -\sin\psi\cos\theta\sin\phi+\cos\psi\cos\phi \ \  & \sin\psi\sin\theta\\
\sin\theta\cos\phi & \sin\theta\sin\phi & \cos\theta
\end{pmatrix}
\end{align}
\end{widetext}
is the rotation matrix where $l_i$, $n_i$, and $m_i$ are the direction cosines and $\phi$, $\theta$, and $\psi$ are the Euler angles. The convention used here is that $\phi$ is the counterclockwise angle around $\hat{z}$ resulting in the transformation $\{\hat{x},\hat{y},\hat{z}\}\to\{\hat{x}^\prime,\hat{y}^\prime,\hat{z}^\prime\}$, $\theta$ is the rotation angle about the $\hat{y}^\prime$ axis resulting in $\{\hat{x}^\prime,\hat{y}^\prime,\hat{z}^\prime\}\to\{\hat{x}^{\prime\prime},\hat{y}^{\prime\prime},\hat{z}^{\prime\prime}\}$, and $\psi$ is the rotation about the $\hat{z}^{\prime\prime}$ axis resulting in $\{\hat{x}^{\prime\prime},\hat{y}^{\prime\prime},\hat{z}^{\prime\prime}\}\to\{\hat{x}^{\prime\prime\prime},\hat{y}^{\prime\prime\prime},\hat{z}^{\prime\prime\prime}\}$. If $A$ is symmetric, the transformation in Eq.~(\ref{eq:matrixtransform}) can be re-expressed as the transformation of a vector with indices running from 1 to 6:
\begin{equation}
A^\prime=\alpha\cdot A ,
\end{equation}
where $A$ is now expressed as $(A_{xx}, A_{yy}, A_{zz}, A_{yz}, A_{xz}, A_{xy})^T$ and 
\begin{widetext}
\begin{align}
\alpha=
\begin{pmatrix}
l_1^2 & m_1^2 & n_1^2 & 2m_1n_1& 2n_1l_1 & 2l_1m_1\\
l_2^2 & m_2^2 & n_2^2 & 2m_2n_2& 2n_2l_2 & 2l_2m_2\\
l_3^2 & m_3^2 & n_3^2 & 2m_3n_3& 2n_3l_3 & 2l_3m_3\\
l_2l_3&m_2m_3&n_2n_3&m_2n_3+n_2m_3&n_2l_3+l_2n_3&l_2m_3+m_2l_3\\
l_1l_3&m_1m_3&n_1n_3&m_1n_3+n_1m_3&n_1l_3+l_1n_3&l_1m_3+m_1l_3\\
l_1l_2&m_1m_2&n_1n_2&m_1n_2+n_1m_2&n_1l_2+l_1n_2&l_1m_2+m_1l_2
\end{pmatrix}.\nonumber
\end{align}
\end{widetext}

A symmetric rank-4 tensor $L$, with indices such as those in Eq.~(\ref{eq:Pi_ab}), that relates two rank 2 tensors $A$ and $B$ through the relation $A=LB$ can be determined in a similar way since $A$ and $B$ transform through the properties outlined above. This leads to the relation $\alpha A=L\alpha B$. It follows that $A=\alpha^{-1}L\alpha B$.
Hence, under a coordinate rotation, $L$ transforms as
\begin{equation}
L^\prime=\alpha^{-1}L\alpha,
\end{equation}
where 
\begin{widetext}
\begin{align}
\alpha^{-1}=
\begin{pmatrix}
l_1^2 & l_2^2 &l_3^2 &2l_2l_3 &2l_1l_3 &2l_1l_2\\
m_1^2 & m_2^2 &m_3^2& 2m_2m_3 & 2m_1m_3& 2m_1m_2\\
n_1^2 & n_2^2 &n_3^2& 2n_2n_3& 2n_1n_3& 2n_1n_2\\
m_1n_1& m_2n_2&m_3n_3& m_2n_3+n_2m_3& m_1n_3+n_1m_3& m_1n_2+n_1m_2\\
n_1l_1& n_2l_2 &n_3l_3& n_2l_3+l_2n_3 &n_1l_3+l_1n_3& n_1l_2+l_1n_2\\
l_1m_1& l_2m_2 &l_3m_3& l_2m_3+m_2l_3& l_1m_3+m_1l_3& l_1m_2+m_1l_2
\end{pmatrix}.
\end{align}
\end{widetext}

Using as an example the 180$^\circ$ rotation about the $z$-axis from Sec.~\ref{sec:visc}, the non-zero elements in $R$ are $l_1=-1$, $m_2=-1$, and $n_3=1$. For this case, 
\begin{align}
\alpha^{-1}=\alpha=
\begin{pmatrix}
1 & 0& 0&0&0&0\\
0&1&0&0&0&0\\
0&0&1&0&0&0\\
0&0&0&-1&0&0\\
0&0&0&0&-1&0\\
0&0&0&0&0&1
\end{pmatrix}.
\end{align}
Application of this operator to Eq.~(\ref{eq:lmatrix}) in Sec.~\ref{sec:visc} leads to many of the conclusions about which elements of the viscosity matrix $L_{ij}$ are zero as a result of the system being invariant when rotating about the magnetic field direction.

\bibliography{viscosity_paper}

\end{document}